\documentclass[aps,pra,twocolumn,showpacs,amsmath,amssymb,floatfix,groupedaddress]{revtex4-1}

\usepackage{graphicx}
\usepackage{bm}
\usepackage{color}

\begin{document}

\title{The significance of the imaginary part of the weak value}

\author{J. Dressel}
\author{A. N. Jordan}
\affiliation{Department of Physics and Astronomy, University of Rochester, Rochester, New York 14627, USA}

\date{\today}


\def\la{\langle}
\def\ra{\rangle}

\newcommand{\op}[1]{\hat{\bm #1}}                
\newcommand{\mean}[1]{\la#1\ra}                  
\newcommand{\cmean}[2]{ { }_{#1}\mean{#2}}       
\newcommand{\ket}[1]{\vert#1\ra}                 
\newcommand{\bra}[1]{\la#1\vert}                 
\newcommand{\ipr}[2]{\la#1\vert#2\ra}            
\newcommand{\opr}[2]{\ket{#1}\bra{#2}}           
\newcommand{\pr}[1]{\opr{#1}{#1}}                
\newcommand{\Tr}[1]{\text{Tr}(#1)}               
\newcommand{\Trd}[1]{\text{Tr}_d(#1)}            
\newcommand{\Trs}[1]{\text{Tr}_s(#1)}            

\begin{abstract}
Unlike the real part of the generalized weak value of an observable, which can in a restricted sense be operationally interpreted as an idealized conditioned average of that observable in the limit of zero measurement disturbance, the imaginary part of the generalized weak value does not provide information pertaining to the observable being measured.  What it does provide is direct information about how the initial state would be unitarily disturbed by the observable operator.  Specifically, we provide an operational interpretation for the imaginary part of the generalized weak value as the logarithmic directional derivative of the post-selection probability along the unitary flow generated by the action of the observable operator.  To obtain this interpretation, we revisit the standard von Neumann measurement protocol for obtaining the real and imaginary parts of the weak value and solve it exactly for arbitrary initial states and post-selections using the quantum operations formalism, which allows us to understand in detail how each part of the generalized weak value arises in the linear response regime.  We also provide exact treatments of qubit measurements and Gaussian detectors as illustrative special cases, and show that the measurement disturbance from a Gaussian detector is purely decohering in the Lindblad sense, which allows the shifts for a Gaussian detector to be completely understood for any coupling strength in terms of a single complex weak value that involves the decohered initial state.
\end{abstract}

\pacs{03.65.Ta,03.65.Ca,03.67.-a}

\maketitle

\section{Introduction}
In their seminal Letter, \citet{Aharonov1988} claimed that they could consistently assign a particular value to an observable that was being weakly measured in a pre- and post-selected ensemble.  To illustrate their technique, they weakly coupled an observable $\op{A}$ to a continuous detector with an initial Gaussian wave-function.  Normally, such a weak von Neumann coupling \cite{VonNeumann1932} would approximately shift the mean of the Gaussian detector wave-function by the expectation value $\bra{\psi_i}\op{A}\ket{\psi_i}$ of $\op{A}$ in the initial state $\ket{\psi_i}$, which would effectively measure $\op{A}$; however, they showed that by post-selecting a final state $\ket{\psi_f}$ after the weak coupling, the mean of the Gaussian detector wave-function could be made to approximately shift by a \emph{complex} quantity that they dubbed the \emph{weak value} of the observable,
\begin{align}\label{eq:aavweakvalue}
  A_w &= \frac{\bra{\psi_f}\op{A}\ket{\psi_i}}{\ipr{\psi_f}{\psi_i}}.
\end{align}
Notably, the weak value expression is not constrained to the eigenvalue range for the observable $\op{A}$, so it can become arbitrarily large for nearly orthogonal pre- and post-selections.

This complex shift in the mean of the Gaussian detector wave-function was only approximate under weak von Neumann coupling and not directly observable, so its significance was not overtly clear; however, the Letter \cite{Aharonov1988} also showed that both the real and imaginary parts of \eqref{eq:aavweakvalue} could be \emph{operationally} obtained from the linear response of the detector under separate conjugate observable measurements.  The practical benefit of this observation was that one could amplify the response of the detector by making a clever choice of post-selection, which potentially allowed for the sensitive determination of other small parameters contributing to the evolution.  

After theoretical clarifications of the derivation in \cite{Duck1989}, experimental confirmation of such amplified detector response soon followed in optical systems \cite{Ritchie1991,Parks1998}.  The amplification has since been used successfully to sensitively measure a variety of phenomena \cite{Hosten2008,Dixon2009,Starling2010,Starling2010b,Turner2011,Hogan2011} to remarkable precision, using both the real and imaginary parts of \eqref{eq:aavweakvalue} as amplification parameters.  Several theoretical extensions of the original derivation of the amplification \cite{Jozsa2007,DiLorenzo2008,Starling2009,Geszti2010,Shikano2010,Cho2010,Wu2011,Haapasalo2011,Parks2011,Zhu2011,Shikano2011,Kofman2011,Nakamura2011,Koike2011,DiLorenzo2011,Pan2011} and several proposals for other amplification measurements have also appeared \cite{Romito2008,Brunner2010,Zilberberg2011,Kagami2011,Li2011}.  In particular, it has been noted that how the amplification effect arises in such a continuous wave-function detector is not intrinsically quantum mechanical, but can also occur in classical wave mechanics \cite{Howell2010}, which has prompted recent study into the mathematical phenomenon of \emph{superoscillations} (e.g. \cite{Aharonov2011,Berry2011}).

Conceptually, however, the weak value expression \eqref{eq:aavweakvalue} has remained quite controversial: since it is generally complex and not constrained to the spectrum of $\op{A}$, how should it be interpreted?  Its primary interpretation in the literature has rested somewhat loosely upon the observation that despite its anomalous behavior one can still decompose an expectation value through the insertion of the identity into an average of weak values, $\bra{\psi_i}\op{A}\ket{\psi_i} = \sum_f |\ipr{\psi_f}{\psi_i}|^2 (\bra{\psi_f}\op{A}\ket{\psi_i}/\ipr{\psi_f}{\psi_i})$, which has the same form as decomposing a classical expectation value $E(X|i)$ into an average of \emph{conditioned expectation values} $E(X|i) = \sum_f P(f|i) E(X|i,f)$.  This observation, together with its approximate appearance operationally in weak conditioned measurements, make it tempting to interpret the weak value as a disturbance-free counter-factual conditioned average that can be assigned to the observable within the context of a pre- and post-selected ensemble even when it is not strictly measured \cite{Aharonov1990,Aharonov2005,Tollaksen2007,Hosoya2011}.

Supporting this point of view is the fact that when the \emph{real part} of \eqref{eq:aavweakvalue} is bounded by the eigenvalue range of $\op{A}$, it agrees with the classical conditioned expectation value for the observable \cite{Aharonov2005}.  Moreover, even when the \emph{real part} is outside the normal eigenvalue range, it still obeys a self-consistent logic \cite{Aharonov2008} and seems to indicate oddly sensible information regarding the operator $\op{A}$.  As such, it has been used quite successfully to analyze and interpret many quantum-mechanical paradoxes both theoretically and experimentally, such as tunneling time \cite{Steinberg1995,Steinberg1995b,Aharonov2002b,Aharonov2003}, vacuum Cherenkov radiation \cite{Rohrlich2002}, cavity QED correlations \cite{Wiseman2002}, double-slit complementarity \cite{Wiseman2003,Mir2007}, superluminal group velocities \cite{Brunner2004}, the N-box paradox \cite{Aharonov1991,Resch2004}, phase singularities \cite{Solli2004}, Hardy's paradox \cite{Aharonov2002,Lundeen2009,Yokota2009,Hosoya2010}, photon arrival time \cite{Wang2006}, Bohmian trajectories \cite{Leavens2005,Wiseman2007,Kocsis2011,Hiley2011}, and Leggett-Garg inequality violations \cite{Williams2008,Goggin2011,Dressel2011}.  

Arguably more important for its status as a quantity pertaining to the measurement of $\op{A}$, however, is the fact that the \emph{real part} of \eqref{eq:aavweakvalue} appears as a stable weak limit point for conditioned measurements even when the detector is not a von Neumann-coupled continuous wave that can experience superoscillatory interference (e.g. \cite{Pryde2005,Goggin2011,Dressel2011,Iinuma2011,Dressel2011b}).  As a result, we can infer that at least the real part of \eqref{eq:aavweakvalue} must have some operational significance specifically pertaining to the measurement of $\op{A}$ that extends beyond the scope of the original derivation.  This observation prompted our Letter \cite{Dressel2010} showing that a principled treatment of a \emph{general conditioned average} of an observable can in fact converge in the weak measurement limit to a generalized expression for the \emph{real part} of \eqref{eq:aavweakvalue},
\begin{align}\label{eq:realweakvalue}
  \text{Re}A_w &= \frac{\Tr{\op{P}_f \{\op{A},\op{\rho}_i\}}}{2\Tr{\op{P}_f \op{\rho}_i}},
\end{align}
where $\{\op{A},\op{\rho}_i\} = \op{A}\op{\rho}_i + \op{\rho}_i\op{A}$ is the \emph{anti-commutator} between the observable operator and an arbitrary initial state $\op{\rho}_i$ represented by a density operator, and where $\op{P}_f$ is an arbitrary post-selection represented by an element from a positive operator-valued measure (POVM).  The general conditioned average converges to \eqref{eq:realweakvalue} provided that the manner in which $\op{A}$ is measured satisfies reasonable sufficiency conditions \cite{Dressel2012,Dressel2011d} that ensure that the disturbance intrinsic to the measurement process does not persist in the weak limit.  

It is in this precise restricted sense that we can operationally interpret the \emph{real part} of the weak value \eqref{eq:realweakvalue} as an \emph{idealized conditioned average of} $\op{A}$ \emph{in the limit of zero measurement disturbance}.  Since it is also the only apparent limiting value of the general conditioned average that no longer depends on \emph{how} the measurement of $\op{A}$ is being made, it is also distinguished as a \emph{measurement context-independent} conditioned average.  These observations provide strong justification for the treatment of the real part of the weak value \eqref{eq:realweakvalue} as a form of value assignment \cite{Aharonov1990,Mermin1993,Spekkens2005,Aharonov2005,Tollaksen2007,Hosoya2011} for the observable $\op{A}$ that depends only upon the preparation and post-selection \footnote{Note that such a value assignment does not violate the Bell-Kochen-Specker theorem \cite{Mermin1993,Spekkens2005,Tollaksen2007} since \eqref{eq:realweakvalue} does not generally obey the product rule, $(AB)_w \neq A_w B_w$.}.

However, we are still left with a mystery: what is the significance of the \emph{imaginary part} of \eqref{eq:aavweakvalue} that appears in the von Neumann measurement, and how does it relate to the operator $\op{A}$?  We can find a partial answer to this question in existing literature (e.g. \cite{Steinberg1995,Steinberg1995b,Aharonov2005,Jozsa2007}) that has associated the appearance of the imaginary part of \eqref{eq:aavweakvalue} in the response of the detector with the intrinsic \emph{disturbance}, or \emph{back-action}, of the measurement process.  For example, regarding continuous von Neumann detectors \citet[p.8]{Aharonov2005} note that ``the imaginary part of the complex weak value can be interpreted as a `bias function' for the posterior sampling point [of the detector].''  Furthermore, they note that ``the weak value of an observable $\op{A}$ is tied to the role of $\op{A}$ as a generator for infinitesimal unitary transformations'' \cite[p.11]{Aharonov2005}.  Similarly, while discussing measurements of tunneling time \citet{Steinberg1995} states that the imaginary part is a ``measure of the back-action on the particle due to the measurement interaction itself'' and that the detector shift corresponding to the imaginary part ``is sensitive to the details of the measurement apparatus (in particular, to the initial uncertainty in momentum), unlike the [shift corresponding to the real part].''  

In this paper, we will augment these observations in the literature by providing a precise operational interpretation of the following generalized expression for the imaginary part of \eqref{eq:aavweakvalue},
\begin{align}\label{eq:imweakvalue}
  \text{Im}A_w &= \frac{\Tr{\op{P}_f (-i[\op{A},\op{\rho}_i])}}{2\Tr{\op{P}_f \op{\rho}_i}},
\end{align}
where $[\op{A},\op{\rho}_i] = \op{A}\op{\rho}_i - \op{\rho}_i\op{A}$ is the \emph{commutator} between $\op{A}$ and the initial state.  We will see that the imaginary part of the weak value does not pertain to the measurement of $\op{A}$ as an observable.  Instead, we will interpret it as half the \emph{logarithmic directional derivative of the post-selection probability along the flow generated by the unitary action of the operator} $\op{A}$.  As such, it provides an explicit measure for the idealized disturbance that the coupling to $\op{A}$ would have induced upon the initial state in the limit that the detector was not measured, which resembles the suggestion by \citet{Steinberg1995}; however, we shall see that the measurement of the detector can strongly alter the state evolution away from that ideal.  The explicit commutator in \eqref{eq:imweakvalue} also indicates that the imaginary part of the weak value involves the operator $\op{A}$ in its role as a generator for unitary transformations as suggested by \citet{Aharonov2005}, in contrast to the real part \eqref{eq:realweakvalue} that involves the operator $\op{A}$ in its role as a measurable observable.

To make it clear how the generalized weak value expressions \eqref{eq:realweakvalue} and \eqref{eq:imweakvalue} and their interpretations arise within a traditional von Neumann detector, we will provide an exact treatment of a von Neumann measurement using the formalism of quantum operations (e.g. \cite{Nielsen2000,Breuer2007,Wiseman2009}).  In addition to augmenting existing derivations in the literature that are concerned largely with understanding the detector response (e.g. \cite{Jozsa2007,DiLorenzo2008,Geszti2010,Wu2011,Haapasalo2011,Parks2011,Zhu2011,Kofman2011,Nakamura2011,Koike2011,DiLorenzo2011,Pan2011}), our exact approach serves to connect the standard treatment of weak values to our more general contextual values analysis that produces the real part \cite{Dressel2010,Dressel2012,Dressel2011d} more explicitly.  We also provide several examples that specialize our exact solution to typically investigated cases: a particular momentum weak value, an arbitrary qubit observable measurement, and a Gaussian detector.  As a consequence, we will show that the Gaussian detector is notable since it induces measurement disturbance that purely \emph{decoheres} the system state into the eigenbasis of $\op{A}$ in the Lindblad sense with increasing measurement strength.  Surprisingly, the pure decoherence allows the shifts in a Gaussian detector to be completely parametrized by a single complex weak value to all orders in the coupling strength, which allows those shifts to be completely understood using our interpretations of that weak value.

The paper is organized as follows.  In \S \ref{sec:vonneumann} we analyze the von Neumann measurement procedure in detail, starting with the traditional unconditioned analysis in \S \ref{sec:traditional}, followed by an operational analysis of the unconditioned case in \S \ref{sec:unconditioned} and the conditioned case in \S \ref{sec:conditioned}.  After obtaining the exact solution for the von Neumann detector response, we consider the weak measurement regime to linear order in the coupling strength in \S \ref{sec:weakvalue}, which clarifies the origins and interpretations of the expressions \eqref{eq:realweakvalue} and \eqref{eq:imweakvalue}.  We discuss the time-symmetric picture in \S \ref{sec:timesymmetry} for completeness.  After a brief Bohmian mechanics example in \S \ref{sec:bohm} that helps to illustrate our interpretation of the weak value, we provide the complete solutions for a qubit observable in \S \ref{sec:qubit} and a Gaussian detector in \S \ref{sec:gaussian}.  Finally, we present our conclusions in \S \ref{sec:conclusion}.

\section{von Neumann Measurement}\label{sec:vonneumann}
The traditional approach for obtaining a complex weak value \cite{Aharonov1988} for a system observable is to post-select a weak Gaussian von Neumann measurement \cite{VonNeumann1932}.  The real and  imaginary parts of the complex weak value then appear as scaled shifts in the conditioned expectations of conjugate detector observables to linear order in the coupling strength.  To clarify how these shifts occur and how the weak value can be interpreted, we shall solve the von Neumann measurement model exactly in the presence of post-selection.

\subsection{Traditional Analysis}\label{sec:traditional}
A von Neumann measurement \cite{VonNeumann1932,Aharonov1988} unitarily couples an operator $\op{A}$ on a \emph{system} Hilbert space $\mathcal{H}_s$ to a momentum operator $\op{p}$ on a continuous \emph{detector} Hilbert space $\mathcal{H}_d$ via a time-dependent interaction Hamiltonian of the form,
\begin{align}\label{eq:hamiltonian}
  \op{H}_I(t) &= g(t) \op{A}\otimes\op{p}.
\end{align}
The interaction profile $g(t)$ is assumed to be a function that is only nonzero over some interaction time interval $t\in[0,T]$.  The interaction is also assumed to be \emph{impulsive} with respect to the natural evolution of the initial joint state $\op{\rho}$ of the system and detector; i.e., the interaction Hamiltonian \eqref{eq:hamiltonian} acts as the total Hamiltonian during the entire interaction time interval.

Solving the Schr\"odinger equation,
\begin{align}\label{eq:schroedinger}
  i\hbar \partial_t \op{U} &= \op{H}_I \op{U},
\end{align}
with the initial condition $\op{U}_0 = \op{1}$ produces a unitary operator,
\begin{align}\label{eq:unitary}
  \op{U}_T &= \exp\left(\frac{g}{i\hbar} \op{A}\otimes\op{p}\right), \\
  g &= \int_0^T dt\, g(t),
\end{align}
that describes the full interaction over the time interval $T$.  The constant $g$ acts as an effective coupling parameter for the impulsive interaction.  If the interaction is \emph{weakly coupled} then $g$ is sufficiently small so that $\op{U}_T \approx \op{1}$ and the effect of the interaction will be approximately negligible; however, we will make no assumptions about the weakness of the coupling \emph{a priori}.

The unitary interaction \eqref{eq:unitary} will entangle the system with the detector so that performing a direct measurement on the detector will lead to an \emph{indirect} measurement being performed on the system.  Specifically, we note that the position operator $\op{x}$ of the detector satisfies the canonical commutation relation $[\op{x},\op{p}]=i\hbar\op{1}_d$, and thus will evolve in the Heisenberg picture of the interaction according to,
\begin{align}\label{eq:xshift}
  (\op{1}_s\otimes\op{x})_T &= \op{U}_T^\dagger (\op{1}_s\otimes\op{x}) \op{U}_T, \\
  &= \op{1}_s\otimes\op{x} + g\op{A}\otimes\op{1}_d. \nonumber 
\end{align}
As a result, measuring the mean of the detector position after the interaction $\mean{x}_T = \Tr{(\op{1}_s\otimes\op{x})_T\op{\rho}}$ will produce,
\begin{align}
  \mean{x}_T &= \mean{x}_0 + g \mean{A}_0.
\end{align}
Hence, the mean of the detector position will be shifted from its initial mean by the mean of the system observable $\op{A}$ in the \emph{initial reduced system state}, linearly scaled by the coupling strength $g$.  For this reason we say that directly measuring the average of the detector position $\op{x}$ results in an indirect measurement of the average of the system observable $\op{A}$.

The detector momentum $\op{p}$, on the other hand, does not evolve in the Heisenberg picture since $[\op{U}_T,\op{1}_s\otimes\op{p}]=0$.  Hence, we expect that measuring the average detector momentum will provide no information about the system observable $\op{A}$.

As discussed in the introduction, however, when one conditions such a von Neumann measurement of the detector upon the outcome of a second measurement made only upon the system, then the conditioned average of \emph{both} the position and the momentum of the detector can experience a shift.  To see why this is so, we will find it useful to switch to the language of \emph{quantum operations} (e.g. \cite{Nielsen2000,Breuer2007,Wiseman2009}) in order to dissect the measurement in more detail.  

\subsection{Quantum Operations}\label{sec:operations}
\subsubsection{Unconditioned Measurement}\label{sec:unconditioned}
As before, we will assume an impulsive interaction in what follows so that any natural time evolution in the joint system and detector state will be negligible on the time scale of the measurement.  (For considerations of the detector dynamics, see \cite{DiLorenzo2008}.)  We will also assume for simplicity of discussion that the initial joint state of the system and detector before the interaction is a product state and that the detector state is pure,
\begin{align}\label{eq:initcond}
  \op{\rho} = \op{\rho}_i \otimes \pr{\psi},
\end{align}
though we will be able to relax this assumption in our final results.  Conceptually, this assumption states that a typical detector will be initially well-calibrated and uncorrelated with the unknown system state that is being probed via the interaction.  

\begin{figure}[t]
  \begin{center}
    \includegraphics[width=3.2in]{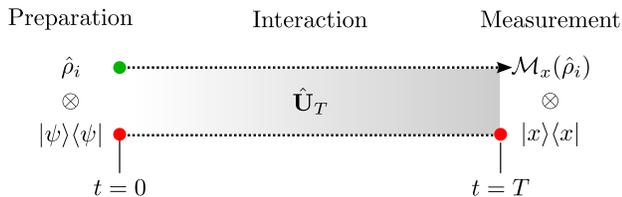}
  \end{center}
  \caption{(color online) Schematic for a von Neumann measurement.  An initially prepared system state $\op{\rho}_i$ and detector state $\pr{\psi}$ become entangled with the von Neumann unitary interaction $\op{U}_T$ \eqref{eq:unitary} over a time interval $T$.  Measuring a particular detector position $x$ after the interaction updates the detector state to $\pr{x}$ and also updates the system state to $\mathcal{M}_x(\op{\rho}_i)$, where $\mathcal{M}_x$ \eqref{eq:measoperx} is an effective measurement operation that encodes the entanglement with and subsequent measurement of the detector. }
  \label{fig:measurement}
\end{figure}

Evolving the initial state with the interaction unitary $\op{U}_T$ \eqref{eq:unitary} will \emph{entangle} the system with the detector.  Hence, subsequently measuring a particular detector position will be equivalent to performing an \emph{operation} $\mathcal{M}_x$ upon the reduced system state, as illustrated in Figure~\ref{fig:measurement},
\begin{align}
  \label{eq:measoperx}
  \mathcal{M}_x(\op{\rho}_i) &= \Trd{(\op{1}_s\otimes\pr{x}) \op{U}_T \op{\rho} \op{U}^\dagger_T} = \op{M}_x \op{\rho}_i \op{M}^\dagger_x, \\ 
  \label{eq:measopx}
  \op{M}_x &= \bra{x}\op{U}_T\ket{\psi}.
\end{align}
where $\Trd{\cdot}$ is the partial trace over the detector Hilbert space, and $\op{M}_x$ is the \emph{Kraus operator} associated with the operation $\mathcal{M}_x$.  Furthermore, since $\ipr{x}{\psi} = \psi(x)$ is the initial detector position wave-function we find $\op{M}_x = \int da\, \exp(-g a \partial_x)\psi(x)\pr{a} = \int da\, \psi(x - g a)\pr{a}$, or, more compactly, $\op{M}_x = \psi(x - g \op{A})$.

If we do \emph{not} perform a subsequent post-selection on the system state, then we trace out the system to find the total probability density for detecting the position $x$,
\begin{align}
  p(x) &= \Trs{\mathcal{M}_x(\op{\rho}_i)} = \Trs{\op{E}_x \op{\rho}_i}, \\
  \label{eq:povmx}
  \op{E}_x &= \op{M}^\dagger_x \op{M}_x = \bra{\psi}\op{U}^\dagger_T(\op{1}_s\otimes\pr{x})\op{U}_T\ket{\psi},
\end{align}
where $\Trs{\cdot}$ is the partial trace over the system Hilbert space.  The \emph{probability operator} $\op{E}_x$ is a positive system operator that encodes the probability of measuring a particular detector position $x$, and can also be written in terms of the initial detector position wave-function as $\op{E}_x = |\psi(x - g\op{A})|^2$.  To conserve probability it satisfies the condition, $\int dx\, \op{E}_x = \op{1}_s$, making the operators $\op{E}_x$ a \emph{positive operator-valued measure} (POVM) on the system space.

Consequently, averaging the position of the detector will effectively average a system observable with the \emph{initial system state},
\begin{align}
  \mean{x}_T &= \int_{-\infty}^\infty dx\, x\, p(x) = \Trs{\op{O} \op{\rho}_i}, \\
  \op{O} &= \int_{-\infty}^\infty dx\, x \op{E}_x = \bra{\psi}\op{U}^\dagger_T(\op{1}_s\otimes\op{x})\op{U}_T\ket{\psi}, \nonumber \\
  &= \mean{x}_0\op{1}_s + g \op{A}, \nonumber
\end{align}
where we see the Heisenberg evolved position operator \eqref{eq:xshift} naturally appear.  

Since the probability operators $\op{E}_x$ are diagonal in the basis of $\op{A}$, then the effective system operator $\op{O}$ will also be diagonal in the same basis.  Hence, by modifying the values that we assign to the position measurements, we can arrange an indirect measurement of any system observable spanned by $\{\op{E}_x\}$ in the basis of $\op{A}$, including $\op{A}$ itself,
\begin{align}\label{eq:cva}
  \op{A} &= \int_{-\infty}^\infty dx\, \left(\frac{x - \mean{x}_0}{g}\right)\, \op{E}_x,
\end{align}
The chosen set of values $(x - \mean{x}_0)/g$ are \emph{contextual values} for $\op{A}$, which can be thought of as a generalized spectrum that relates $\op{A}$ to the specific POVM $\{\op{E}_x\}$ associated with the \emph{measurement context} $\{\mathcal{M}_x\}$ \cite{Dressel2010,Dressel2012,Dressel2011d}.  They are not the only values that we could assign to the position measurement in order to obtain the equality \eqref{eq:cva}, but they are arguably the simplest to obtain and compute, as well as the most frequently used in the literature.  It is in this precise sense that we can say that the von Neumann coupling leads to an indirect measurement of the average of $\op{A}$ in the absence of post-selection.

The measurement of $\op{A}$ comes at a cost, however, since the system state is necessarily \emph{disturbed} by the operations $\mathcal{M}_x$ in order to obtain the probability operators $\op{E}_x$.  The state may even be disturbed more than is strictly required to make the measurement of $\op{A}$, which can be seen by rewriting the measurement operators in polar form, $\op{M}_x = \op{U}_x |\op{E}_x|^{1/2}$, with the positive root of the probability operator $|\op{E}_x|^{1/2}$ and an additional unitary operator $\op{U}_x$.  This decomposition implies that $\mathcal{M}_x$ splits into an effective composition of two distinct operations,
\begin{subequations}\label{eq:extraunitary}
\begin{align}
  \mathcal{M}_x(\op{\rho}_i) &= \mathcal{U}_x(\mathcal{E}_x(\op{\rho}_i)), \\
  \mathcal{E}_x(\op{\rho}_i) &= |\op{E}_x|^{1/2}\op{\rho}_i|\op{E}_x|^{1/2}, \\
  \mathcal{U}_x(\op{\rho}'_i) &= \op{U}_x\op{\rho}'_i\op{U}^\dagger_x.
\end{align}
\end{subequations}
We can interpret the operation $\mathcal{E}_x$ that involves only the roots of the probability operator $|\op{E}_x|^{1/2}$ as the \emph{pure measurement operation} producing $\op{E}_x$.  That is, it represents the \emph{minimum necessary disturbance} that one must make to the initial state in order to extract a measurable probability.  The second operation $\mathcal{U}_x$ unitarily disturbs the initial state, but does not contribute to $\op{E}_x$.  Since only $\op{E}_x$ can be used to infer information about $\op{A}$ through the identity \eqref{eq:cva}, we conclude that the disturbance from $\mathcal{U}_x$ is superfluous.

To identify the condition for eliminating $\mathcal{U}_x$, we can rewrite the Kraus operator \eqref{eq:measopx} using the polar form of the initial detector position wave-function $\psi(x) = \exp(i\psi_s(x))\psi_r(x)$,
\begin{align}\label{eq:polarmx}
  \op{M}_x &= \exp(i\psi_s(x - g\op{A})) \psi_r(x - g\op{A}).
\end{align}
The phase factor becomes the unitary operator $\op{U}_x = \exp(i\psi_s(x - g \op{A}))$ for $\mathcal{U}_x$, while the magnitude becomes the required positive root $|\op{E}_x|^{1/2} = \psi_r(x - g\op{A})$ for $\mathcal{E}_x$.  Hence, to eliminate the superfluous operation $\mathcal{U}_x$ from a von Neumann measurement with coupling Hamiltonian \eqref{eq:hamiltonian}, one must use a \emph{purely real} initial detector wave-function in position.

For contrast, measuring only a particular detector momentum $p$ will be equivalent to performing a different operation $\mathcal{N}_p$ upon the reduced system state,
\begin{align}
  \mathcal{N}_p(\op{\rho}_i) &= \Trd{(\op{1}_s\otimes\pr{p}) \op{U}_T \op{\rho} \op{U}_T^\dagger} = \op{N}_p \op{\rho}_i \op{N}^\dagger_p, \\
  \label{eq:measopp}
  \op{N}_p &= \bra{p}\op{U}_T\ket{\psi} = \exp\left(\frac{gp}{i\hbar} \op{A}\right)\ipr{p}{\psi}. 
\end{align}
The Kraus operator $\op{N}_p$ has a purely unitary factor containing $\op{A}$ that will disturb the system, regardless of the form of the initial momentum wave-function $\ipr{p}{\psi}$.  Moreover, the probability operator associated with the momentum measurement has the form,
\begin{align}
  \label{eq:povmp}
  \op{F}_p &= \op{N}^\dagger_p \op{N}_p = |\ipr{p}{\psi}|^2 \op{1}_s,
\end{align}
which can only be used to measure the identity $\op{1}_s$.

For completeness we also briefly note that the conjugate Kraus operators $\op{M}_x$ and $\op{N}_p$ are related through a Fourier transform,
\begin{subequations}
\begin{align}
  \op{N}_p &= \frac{1}{\sqrt{2\pi\hbar}}\int_{-\infty}^\infty dx\, e^{-ipx/\hbar} \op{M}_x, \\
  \op{M}_x &= \frac{1}{\sqrt{2\pi\hbar}}\int_{-\infty}^\infty dp\, e^{ipx/\hbar} \op{N}_p,
\end{align}
\end{subequations}
and that both detector probability operators can be obtained as marginals of a \emph{Wigner quasi-probability operator} on the system Hilbert space,
\begin{subequations}
\begin{align}
  \label{eq:wigner}
  \op{W}_{x,p} &= \frac{1}{\pi\hbar}\int_{-\infty}^\infty dy\, e^{2ipy/\hbar} \op{M}^\dagger_{x+y}\op{M}_{x-y}, \\
  \op{E}_x &= \int_{-\infty}^\infty dp\, \op{W}_{x,p}, \\
  \op{F}_p &= \int_{-\infty}^\infty dx\, \op{W}_{x,p}.
\end{align}
\end{subequations}
In the absence of interaction, then the Wigner quasi-probability operator reduces to the Wigner quasi-probability distribution $W(x,p)$ for the initial \emph{detector} state, $\op{W}_{x,p} \xrightarrow{g=0} W(x,p) \op{1}_s$.

\subsubsection{Conditioned Measurement}\label{sec:conditioned}
\begin{figure}[t]
  \begin{center}
    \includegraphics[width=2in]{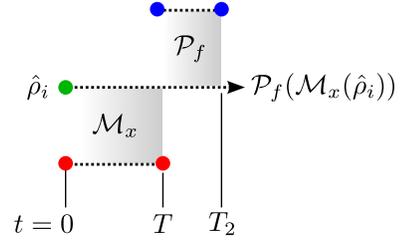}
  \end{center}
  \caption{(color online) Schematic for a sequence of two indirect measurements.  After the von Neumann interaction and measurement of $x$ illustrated in Figure~\ref{fig:measurement} that produces the effective measurement operation $\mathcal{M}_x$ upon the initial system state, a second detector interacts impulsively with the system for a time interval $T_2 - T$.  The second detector is then measured to have a particular outcome $f$, which updates the system state to $\mathcal{P}_f(\mathcal{M}_x(\op{\rho}_i))$, where $\mathcal{P}_f$ is another measurement operation.  Taking the trace of the final system state will then produce the joint probability densities \eqref{eq:jointprobs}.}
  \label{fig:measurement2}
\end{figure}

To post-select the system, an experimenter must perform a second measurement after the von Neumann measurement and filter the two-measurement event space based on the outcomes for the second measurement.  In other words, the experimenter keeps only those pairs of outcomes for which the second outcome satisfies some constraint.  The remaining measurement pairs can then be averaged to produce \emph{conditioned averages} of the first measurement.

If we represent the second measurement as a set of probability operators $\{\op{P}_f\}$ indexed by some parameter $f$ that can be derived analogously to \eqref{eq:povmx} from a set of operations $\{\mathcal{P}_f\}$ as illustrated in Figure~\ref{fig:measurement2}, then the total joint probability densities for the ordered sequences of measurement outcomes $(x,f)$ and $(p,f)$ will be,
\begin{subequations}\label{eq:jointprobs}
\begin{align}
  p(x,f) &= \Trs{\op{P}_f \mathcal{M}_x(\op{\rho}_i)} = \Trs{\op{E}_{x,f}\op{\rho}_i}, \\
  p(p,f) &= \Trs{\op{P}_f \mathcal{N}_p(\op{\rho}_i)} = \Trs{\op{F}_{p,f}\op{\rho}_i}, 
\end{align}
\end{subequations}
where the joint probability operators,
\begin{subequations}\label{eq:jointpovm}
\begin{align}
  \op{E}_{x,f} &= \op{M}_x^\dagger\op{P}_f\op{M}_x, \\
  \op{F}_{p,f} &= \op{N}_p^\dagger\op{P}_f\op{N}_p.
\end{align}
\end{subequations}
are not simple products of the post-selection $\op{P}_f$ and the probability operators \eqref{eq:povmx} or \eqref{eq:povmp}.  Those operators can be recovered, however, by marginalizing over the index $f$, since the post-selection probability operators must satisfy a POVM condition $\sum_f \op{P}_f = \op{1}_s$.  

The joint probabilities \eqref{eq:jointprobs} will contain information not only about the first measurement and the initial system state, but also about the second measurement and any \emph{disturbance} to the initial state that occurred due to the first measurement.  In particular, the joint probability operators \eqref{eq:jointpovm} can no longer satisfy the identity \eqref{eq:cva} due to the second measurement, so averaging the probabilities \eqref{eq:jointprobs} must reveal more information about the measurement process than can be obtained solely from the operator $\op{A}$, the initial state $\op{\rho}_i$, and the post-selection $\op{P}_f$.  As a poignant example, the unitary disturbance $\mathcal{U}_x$ in \eqref{eq:extraunitary} that did not contribute to the operator identity \eqref{eq:cva} \emph{will} contribute to the joint probability operators, $\op{E}_{x,f} = |\op{E}_x|^{1/2}\op{U}_x^\dagger\op{P}_f\op{U}_x|\op{E}_x|^{1/2}$.

The total probability for obtaining the post-selection outcome $f$ can be obtained by marginalizing over either $x$ or $p$ in the joint probabilities,
\begin{align}
  p(f) &= \int_{-\infty}^\infty dx\, p(x,f) = \int_{-\infty}^\infty dp\, p(p,f), \\
  &= \Trs{\op{P}_f \mathcal{E}(\op{\rho}_i)}, \nonumber \\
  \label{eq:oper}
  \mathcal{E}(\op{\rho}_i) &= \Trd{\op{U}_T(\op{\rho}_i\otimes\pr{\psi})\op{U}^\dagger_T},
\end{align}
where the operation $\mathcal{E}$ is the total \emph{non-selective} measurement that has been performed on $\op{\rho}_i$.  Since $\mathcal{E}$ is not the identity operation, it represents the total \emph{disturbance} intrinsic to the measurement process.  It includes unitary evolution of the reduced system state due to the interaction Hamiltonian \eqref{eq:hamiltonian}, as well as \emph{decoherence} stemming from entanglement with the measured detector.

By experimentally filtering the event pairs to keep only a particular outcome $f$ of the second measurement, an experimenter can obtain the conditional probabilities,
\begin{subequations}
\begin{align}
  p(x|f) &= \frac{p(x,f)}{p(f)} = \frac{\Trs{\op{P}_f \mathcal{M}_x(\op{\rho}_i)}}{\Trs{\op{P}_f \mathcal{E}(\op{\rho}_i)}}, \\
  p(p|f) &= \frac{p(p,f)}{p(f)} = \frac{\Trs{\op{P}_f \mathcal{M}_p(\op{\rho}_i)}}{\Trs{\op{P}_f \mathcal{E}(\op{\rho}_i)}}, 
\end{align}
\end{subequations}
which can then be averaged to find the exact conditioned averages for the detector position and momentum,
\begin{subequations}\label{eq:condavxp}
\begin{align}
  \cmean{f}{x}_T &= \int_{-\infty}^\infty dx\, x\, p(x|f) = \frac{\Trs{\op{P}_f \mathcal{X}_T(\op{\rho}_i)}}{\Trs{\op{P}_f \mathcal{E}(\op{\rho}_i)}}, \\
  \cmean{f}{p}_T &= \int_{-\infty}^\infty dp\, p\, p(p|f) = \frac{\Trs{\op{P}_f \mathcal{P}_T(\op{\rho}_i)}}{\Trs{\op{P}_f \mathcal{E}(\op{\rho}_i)}},
\end{align}
\end{subequations}
where,
\begin{subequations}\label{eq:operxpT}
\begin{align}
  \label{eq:operx}
  \mathcal{X}_T(\op{\rho}_i) &= \Trd{(\op{1}_s\otimes\op{x})\op{U}_T(\op{\rho}_i\otimes\pr{\psi})\op{U}^\dagger_T}, \\
  \label{eq:operp}
  \mathcal{P}_T(\op{\rho}_i) &= \Trd{(\op{1}_s\otimes\op{p})\op{U}_T(\op{\rho}_i\otimes\pr{\psi})\op{U}^\dagger_T},
\end{align}
\end{subequations}
are detector averaging operations that affect the system state before the measurement of the post-selection is performed.  It is worth noting at this point that we can relax the assumption \eqref{eq:initcond} made about the initial state in the exact operational expressions \eqref{eq:oper} and \eqref{eq:operxpT}.  Similarly, if different contextual values are used to average the conditional probabilities in \eqref{eq:condavxp}, then corresponding detector observables with the same spectra will appear in the operations \eqref{eq:operxpT} in place of $\op{x}$ or $\op{p}$; for example, averaging the values $\alpha(x) = (x-\mean{x}_0)/g$ used in \eqref{eq:cva} will replace the detector observable $\op{x}$ in \eqref{eq:operxpT} with $\op{\alpha} = \int_{-\infty}^\infty dx\, \alpha(x)\, \pr{x}$.

To better interpret \eqref{eq:operxpT}, we bring the detector operators inside the unitary operators in \eqref{eq:operxpT} using the canonical commutation relations as in \eqref{eq:xshift},
\begin{subequations}\label{eq:operxpA}
\begin{align}
  \mathcal{X}_T(\op{\rho}_i) &= \mathcal{X}(\op{\rho}_i) + g\,\mathcal{E}(\{\op{A},\op{\rho}_i\}/2), \\
  \mathcal{P}_T(\op{\rho}_i) &= \mathcal{P}(\op{\rho}_i),
\end{align}
\end{subequations}
which splits the $\mathcal{X}_T$ operation into two operations but only changes the form of $\mathcal{P}_T$.  The operation proportional to $g$ disturbs the symmetrized product $\{\op{A},\op{\rho}_i\}/2 = (\op{A}\op{\rho}_i + \op{\rho}_i\op{A})/2$ of the \emph{initial system state} with the operator $\op{A}$, while the operations,
\begin{subequations}\label{eq:operxp}
\begin{align}
  \mathcal{X}(\op{\rho}_i) &= \Trd{\op{U}_T(\op{\rho}_i\otimes\{\op{x},\pr{\psi}\}/2)\op{U}^\dagger_T}, \\
  \mathcal{P}(\op{\rho}_i) &= \Trd{\op{U}_T(\op{\rho}_i\otimes\{\op{p},\pr{\psi}\}/2)\op{U}^\dagger_T},
\end{align}
\end{subequations}
disturb the symmetrized products of the \emph{initial detector state} with the detector operators.

The form of the equations \eqref{eq:operxpA} clearly illustrates how the post-selection will affect the measurement.  If the post-selection is the identity operator, $\op{P}_f = \op{1}_s$, then the unitary operators $\op{U}_T$ causing the total disturbance of the initial state will cancel through the cyclic property of the total trace in \eqref{eq:condavxp}, leaving the averages in the \emph{initial} states that were previously obtained,
\begin{subequations}\label{eq:detectormeans}
\begin{align}
  \mean{x}_T &= \mean{x}_0 + g\mean{A}_0, \\
  \mean{p}_T &= \mean{p}_0.
\end{align}
\end{subequations}
In this sense, commuting the detector operators $\op{x}$ and $\op{p}$ in \eqref{eq:operxpT} through the unitary operators to arrive at \eqref{eq:operxpA} is equivalent to evolving them in the Heisenberg picture back from the time of measurement $T$ to the initial time $0$ in order to compare them with the initial states.  However, the presence of the post-selection operator $\op{P}_f$ will now generally spoil the cancelation of the unitary operators that is implicit in the Heisenberg picture, leading to corrections from the disturbance between the pre- and post-selection.  

The symmetrized products in \eqref{eq:operxpA} indicate the measurement being made on the \emph{initial} states of the system and detector, which is then further disturbed by the unitary operators $\op{U}_T$ as a consequence of the coupling Hamiltonian \eqref{eq:hamiltonian}.  The post-selection both conditions those measurements and reveals the disturbance, which corrects each term in \eqref{eq:detectormeans}, yielding the final exact expressions,
\begin{subequations}\label{eq:corrections}
\begin{align}
  \label{eq:xcorr}
  \cmean{f}{x}_T &= \frac{\Trs{\op{P}_f \mathcal{X}(\op{\rho}_i)}}{\Trs{\op{P}_f \mathcal{E}(\op{\rho}_i)}} + g \frac{\Trs{\op{P}_f \mathcal{E}(\{\op{A},\op{\rho}_i\})}}{2\Trs{\op{P}_f \mathcal{E}(\op{\rho}_i)}}, \\
  \label{eq:pcorr}
  \cmean{f}{p}_T &= \frac{\Trs{\op{P}_f \mathcal{P}(\op{\rho}_i)}}{\Trs{\op{P}_f \mathcal{E}(\op{\rho}_i)}}.
\end{align}
\end{subequations}

\section{The Weak Value}\label{sec:weakvalue}
If it were possible to leave the system state undisturbed while still allowing the measurement of $\op{A}$, then we would na\"{i}vely expect the disturbance $\mathcal{E}$ to reduce to the identity operation.  Similarly, we would na\"{i}vely expect the operations $\mathcal{X}$ and $\mathcal{P}$ would reduce to $\mean{x}_0$ and $\mean{p}_0$ multiplying the identity operation, respectively.  As a result, the conditioned averages \eqref{eq:corrections} would differ from the unconditioned averages \eqref{eq:detectormeans} solely by the replacement of the average $\mean{A}_0$ with the \emph{real part} \eqref{eq:realweakvalue} of the complex \emph{generalized weak value} expression,
\begin{align}\label{eq:weakvalue}
  A_w &= \frac{\Trs{\op{P}_f \op{A}\op{\rho}_i}}{\Trs{\op{P}_f \op{\rho}_i}}.
\end{align}
Since this expression depends solely upon the initial state $\op{\rho}_i$, the post-selection $\op{P}_f$, and the operator $\op{A}$, we are na\"{i}vely tempted to give $\text{Re}A_w$ an intuitive interpretation as the \emph{ideal conditioned expectation} of $\op{A}$ in a pre- and post-selected state with no intermediate measurement disturbance.  However, it is strictly impossible to remove the disturbance from the measurement while still making the measurement, so we cannot rely on this sort of reasoning.  We can make a similar interpretation in a restricted sense, however, by making the coupling strength $g$ sufficiently small to reduce the disturbance to a minimal amount that still allows the measurement to be made.  

To see how the operations $\mathcal{E}$, $\mathcal{X}$, and $\mathcal{P}$ in \eqref{eq:oper} and \eqref{eq:operxp} depend on the coupling strength $g$, we expand them perturbatively,
\begin{subequations}\label{eq:opexpand}
\begin{align}
  \mathcal{E}(\op{\rho}_i) &= \sum_{n=0} \frac{1}{n!} \left(\frac{g}{i\hbar}\right)^n \mean{p^n}_0 (\text{ad}\op{A})^n(\op{\rho}_i), \\
  \mathcal{X}(\op{\rho}_i) &= \sum_{n=0} \frac{1}{n!} \left(\frac{g}{i\hbar}\right)^n \frac{\mean{\{p^n,x\}}_0}{2} (\text{ad}\op{A})^n(\op{\rho}_i), \\
  \mathcal{P}(\op{\rho}_i) &= \sum_{n=0} \frac{1}{n!} \left(\frac{g}{i\hbar}\right)^n \mean{p^{n+1}}_0 (\text{ad}\op{A})^n(\op{\rho}_i),
\end{align}
\end{subequations}
where the operation $(\text{ad}\op{A})(\cdot) = [\op{A},\cdot]$ is the left action of $\op{A}$ in the adjoint representation of its Lie algebra, which takes the form of a commutator.  That is, $(\text{ad}\op{A})$ explicitly describes how $\op{A}$ \emph{disturbs} the initial state due to the interaction that measures it.  

The initial detector state plays a critical role in \eqref{eq:opexpand} by determining the various moments, $\mean{p^n}_0$, $\mean{p^{n+1}}_0$ and $\mean{\{p^n,x\}/2}_0$ that appear in the series expansions.  Notably, if we make the initial detector wave-function purely real so that it minimally disturbs the system state then all moments containing odd powers of $\op{p}$ will vanish.  We conclude that those moments of the disturbance operations are superfluous for obtaining the measurable probabilities that allow the measurement of $\op{A}$, while the moments with even powers of $\op{p}$ are necessary.

After expanding the corrections \eqref{eq:corrections} to first order in $g$, we obtain the linear response of the conditioned detector means due to the interaction,
\begin{subequations}\label{eq:linearraw}
\begin{align}
  \cmean{f}{x}_T &\to \mean{x}_0 + \frac{g}{i\hbar}\frac{\mean{\{p,x\}}_0}{2} \frac{\Trs{\op{P}_f (\text{ad}\op{A})(\op{\rho}_i)}}{\Trs{\op{P}_f \op{\rho}_i}} \\
  &\qquad + g \frac{\Trs{\op{P}_f\{\op{A}, \op{\rho}_i\}}}{2\Trs{\op{P}_f \op{\rho}_i}}, \nonumber \\
  \cmean{f}{p}_T &\to \mean{p}_0 + \frac{g}{i\hbar}\mean{p^2}_0 \frac{\Trs{\op{P}_f (\text{ad}\op{A})(\op{\rho}_i)}}{\Trs{\op{P}_f \op{\rho}_i}}.
\end{align}
\end{subequations}
Measurements for which this linear response is a good approximation are known as \emph{weak measurements}.  

After introducing the complex generalized weak value \eqref{eq:weakvalue}, we can write the linear response formulas in a more compact form,
\begin{subequations}\label{eq:linear}
\begin{align}
  \cmean{f}{x}_T &= \mean{x}_0 + \frac{g}{\hbar}\frac{\mean{\{p,x\}}_0}{2}\, (2 \text{Im}A_w) + g\, \text{Re}A_w, \\
  \cmean{f}{p}_T &= \mean{p}_0 + \frac{g}{\hbar}\mean{p^2}_0\, (2 \text{Im}A_w),
\end{align}
\end{subequations}
in terms of not only its real part, but also \emph{twice its imaginary part}.

If the initial detector position wave-function $\psi(x)$ is purely real, so that the measurement is minimally disturbing, then $\mean{\{p,x\}/2}_0$ will vanish, leaving only $\text{Re}A_w$ in $\cmean{f}{x}_T$ as we na\"{i}vely reasoned before.  However, the term proportional to $2 \text{Im}A_w$ will not vanish in $\cmean{f}{p}_T$ to linear order in $g$, making it an element of measurement disturbance that persists even for minimally disturbing weak measurements.

These linear response formulas for the von Neumann measurement have also been obtained and discussed in the literature with varying degrees of generality and rigor (e.g. \cite{Aharonov1988,Duck1989,Jozsa2007,Aharonov2008,DiLorenzo2008,Geszti2010,Wu2011,Haapasalo2011,Parks2011,Kofman2011,Nakamura2011,Koike2011,DiLorenzo2011,Pan2011}).  However, our derivation has a conceptual advantage in that we see explicitly how the origins of the real and imaginary parts of the weak value differ with respect to the measurement of $\op{A}$.  We are therefore in a position to give concrete interpretations for each part.

The real part \eqref{eq:realweakvalue} of the weak value $\text{Re}A_w$ stems directly from the part of the conditioned shift of the detector pointer that corresponds to the measurement of $\op{A}$ and does not contain any further perturbation induced by the measurement coupling that would be indicated by factors of $(\text{ad}\op{A})$.  As a result, it can be interpreted as an idealized limit point for the average of $\op{A}$ in the initial state $\op{\rho}_i$ that has been conditioned on the post-selection $\op{P}_f$ without any appreciable intermediate measurement disturbance.  To support this point of view, we have also shown in \cite{Dressel2010,Dressel2012,Dressel2011d} that $\text{Re}A_w$ appears naturally as such a limit point for minimally disturbing measurements that are not of von Neumann type, provided that those measurements satisfy reasonable sufficiency conditions regarding the measurability of $\op{A}$.

The imaginary part \eqref{eq:imweakvalue} of the weak value $\text{Im}A_w$, on the other hand, stems directly from the \emph{disturbance} of the measurement and explicitly contains $(\text{ad}\op{A})$, which is the action of $\op{A}$ as a generator for unitary evolution due to the specific Hamiltonian \eqref{eq:hamiltonian}.  The factor $2\text{Im}A_w$ appears in \eqref{eq:linear} along with information about the initial detector momentum that is being coupled to $\op{A}$ in the Hamiltonian \eqref{eq:hamiltonian}, as well as factors of $\hbar$, in stark contrast to the real part.  How then can it be interpreted?

The significance of $2\text{Im}A_w$ becomes more clear once we identity the \emph{directional derivative operation} that appears in its numerator,
\begin{align}\label{eq:flowa}
  \delta_A(\cdot) &= -i(\text{ad}\op{A})(\cdot).
\end{align}
That is, $\delta_A(\op{\rho}_i)$ indicates the \emph{rate of change} of the initial state $\op{\rho}_i$ along a \emph{flow} in state-space generated by $\op{A}$.  

The directional derivative should be familiar from the Schr\"{o}dinger equation written in the form $\partial_t \op{\rho} = [\op{H},\op{\rho}]/i\hbar = \delta_\Omega(\op{\rho})$, where the scaled Hamiltonian $\op{\Omega} = \op{H}/\hbar$ is a characteristic frequency operator.  The integration of this equation is a unitary operation in exponential form $\op{\rho}(t) = \exp(t\delta_\Omega)(\op{\rho}(0)) = \exp(-it\op{\Omega})\op{\rho}(0)\exp(it\op{\Omega})$ that specifies a \emph{flow} in state space, which is a collection of curves that is parametrized both by a time parameter $t$ and by the initial condition $\op{\rho}(0)$.  Specifying the initial condition $\op{\rho}(0) = \op{\rho}_i$, picks out the specific curve from the flow that contains $\op{\rho}_i$.  The directional derivative of the initial state along that specific curve is then defined in the standard way, $\partial_t \op{\rho}(t)|_{t=0} = \delta_\Omega(\op{\rho}_i)$.

The fact that the quantum state space is always a continuous manifold of states allows such a flow to be defined in a similar fashion using any Hermitian operator, such as $\op{A}$, as a generator.  Analogously to time evolution, such a flow has the form of a unitary operation, $\op{\rho}(\varepsilon) = \exp(\varepsilon\delta_A)(\op{\rho}(0))$, where the real parameter $\varepsilon$ for the flow has units inverse to $\op{A}$.  Therefore, taking the directional derivative of $\op{\rho}_i$ along the specific curve of this flow that passes through $\op{\rho}_i$ will produce \eqref{eq:flowa}.  For an explicit example that we will detail in \S\ref{sec:qubit}, the state-space of a qubit can be parametrized as the continuous volume of points inside the unit Bloch sphere; the derivative \eqref{eq:flowa} produces the vector field illustrated in Figure~\ref{fig:blochfield} tangent to the flow corresponding to Rabi oscillations of the qubit.

With this intuition in mind, we define the post-selection probability for measuring $\op{P}_f$ given an initial state $\op{\rho}_i(\varepsilon)=\exp(\varepsilon\delta_A)(\op{\rho}_i)$ that is changing along the flow generated by $\op{A}$,
\begin{align}
  p_f(\varepsilon) &= \Trs{\op{P}_f \op{\rho}_i(\varepsilon)}.
\end{align}
The logarithmic directional derivative of this post-selection probability then produces the factor $2\text{Im}A_w$ that appears in \eqref{eq:linear},
\begin{align}\label{eq:imweakvaluedef}
  2\text{Im}A_w &= \partial_\varepsilon \ln p_f(\varepsilon) \big|_{\varepsilon=0},
\end{align}
which is our main result.

In words, \emph{the imaginary part of the weak value is half the logarithmic directional derivative of the post-selection probability along the natural unitary flow generated by} $\op{A}$.  It does not provide any information about the measurement of $\op{A}$ as an observable, but rather \emph{indicates an instantaneous exponential rate of change in the post-selection probability due to disturbance of the initial state caused by} $\op{A}$ \emph{in its role as a generator for unitary transformations}.  Specifically, for small $\varepsilon$ we have the approximate relation,
\begin{align}
  p_f(\varepsilon) &\approx p_f(0)(1 + (2\text{Im}A_w)\varepsilon).
\end{align} 
For a pure initial state $\op{\rho}_i = \pr{\psi_i}$ and a projective post-selection $\op{P}_f = \pr{\psi_f}$, the expression \eqref{eq:imweakvaluedef} simplifies,
\begin{align}
  2\text{Im}A_w &= \partial_\varepsilon \ln |\bra{\psi_f}\exp(-i\varepsilon\op{A})\ket{\psi_i}|^2 \big|_{\varepsilon=0}.
\end{align}

Hence, the corrections containing $2\text{Im}A_w$ that appear in \eqref{eq:linear} stem directly from how the specific von Neumann Hamiltonian \eqref{eq:hamiltonian} unitarily disturbs the initial system state infinitesimally prior to any additional disturbance induced by the measurement of the detector.  Conceptually, the coupling induces a natural unitary flow of the initial system state generated by $\op{A}$, which for infinitesimal $g$ changes the joint probability $p(p,f)$ for a specific $p$ by the amount $(2\text{Im}A_w)(g p/\hbar)$, where $gp/\hbar$ is the infinitesimal parameter $\varepsilon$ in \eqref{eq:imweakvaluedef} that has units inverse to $\op{A}$.  Averaging this correction to the joint probability with the detector observables $\op{x}$ or $\op{p}$ produces the correction terms in \eqref{eq:linear}.

\section{Time Symmetry}\label{sec:timesymmetry}
As noted in \cite{Aharonov1964,Aharonov2008}, a quantum system that has been pre- and post-selected exhibits time symmetry.  We can make the time symmetry more apparent in our operational treatment by introducing the \emph{retrodictive state},
\begin{align}\label{eq:retrodictivef}
  \op{\rho}_f &= \op{P}_f / \Tr{\op{P}_f},
\end{align}
associated with the post-selection (see, e.g., \cite{Pegg2002,Amri2011}) and rewriting our main results in the time-reversed retrodictive picture.

After cancelling normalization factors, the detector response \eqref{eq:corrections} for a system retrodictively prepared in the final state $\op{\rho}_f$ that has been conditioned on the \emph{pre-selection} measurement producing the initial system state $\op{\rho}_i$ has the form, 
\begin{subequations}\label{eq:retrocorrections}
\begin{align}
  \label{eq:retroxcorr}
  \cmean{i}{x}_T &= \frac{\Trs{\mathcal{X}^*(\op{\rho}_f) \op{\rho}_i}}{\Trs{\mathcal{E}^*(\op{\rho}_f) \op{\rho}_i}} + g \frac{\Trs{\{\mathcal{E}^*(\op{\rho}_f),\op{A}\} \op{\rho}_i}}{2\Trs{\mathcal{E}^*(\op{\rho}_f)\op{\rho}_i}}, \\
  \label{eq:retropcorr}
  \cmean{i}{p}_T &= \frac{\Trs{\mathcal{P}^*(\op{\rho}_f) \op{\rho}_i}}{\Trs{\mathcal{E}^*(\op{\rho}_f) \op{\rho}_i}},
\end{align}
\end{subequations}
where the retrodictive operations $\mathcal{E}^*$, $\mathcal{X}^*$, and $\mathcal{P}^*$ are the adjoints of the predictive operations in \eqref{eq:oper} and \eqref{eq:operxp},
\begin{subequations}\label{eq:retroops}
\begin{align}
  \mathcal{E}^*(\op{\rho}_f) &= \bra{\psi}\op{U}^\dagger_T\op{\rho}_f\op{U}_T\ket{\psi}, \\
  \mathcal{X}^*(\op{\rho}_f) &= \bra{\psi}\{\op{x},\op{U}^\dagger_T\op{\rho}_f\op{U}_T\}/2\ket{\psi}, \\
  \mathcal{P}^*(\op{\rho}_f) &= \bra{\psi}\{\op{p},\op{U}^\dagger_T\op{\rho}_f\op{U}_T\}/2\ket{\psi}.
\end{align}
\end{subequations}

Notably, the symmetric product with $\op{A}$ that appears in the detector response \eqref{eq:retrocorrections} involves the retrodictive state that has been evolved back to the initial time due to the non-selective measurement operation $\mathcal{E}^*$.  Hence, in both pictures the measurement of $\op{A}$ is being made with respect to the same initial time.  

After expanding the retrodictive operations perturbatively as in \eqref{eq:opexpand},
\begin{subequations}\label{eq:retroopexpand}
\begin{align}
  \mathcal{E}^*(\op{\rho}_f) &= \sum_{n=0} \frac{1}{n!} \left(\frac{g}{i\hbar}\right)^n \mean{p^n}_0 (\text{ad}^*\op{A})^n(\op{\rho}_f), \\
  \mathcal{X}^*(\op{\rho}_f) &= \sum_{n=0} \frac{1}{n!} \left(\frac{g}{i\hbar}\right)^n \frac{\mean{\{p^n,x\}}_0}{2} (\text{ad}^*\op{A})^n(\op{\rho}_f), \\
  \mathcal{P}^*(\op{\rho}_f) &= \sum_{n=0} \frac{1}{n!} \left(\frac{g}{i\hbar}\right)^n \mean{p^{n+1}}_0 (\text{ad}^*\op{A})^n(\op{\rho}_f),
\end{align}
\end{subequations}
where $(\text{ad}^*\op{A})(\cdot) = -(\text{ad}\op{A})(\cdot) = [\cdot,\op{A}]$ is the right action of $\op{A}$, then the linear response of the detector \eqref{eq:linear} can be written in terms of the retrodictive forms of the real and imaginary parts of the complex weak value,
\begin{align}\label{eq:retroweakvalue}
  \text{Re}A_w &= \frac{\Trs{\{\op{\rho}_f,\op{A}\} \op{\rho}_i}}{2\Trs{\op{\rho}_f \op{\rho}_i}}, \\
  2\text{Im}A_w &= \frac{\Trs{(-i[\op{\rho}_f,\op{A}]) \op{\rho}_i}}{\Trs{\op{\rho}_f \op{\rho}_i}}, \\
  &= \partial_\varepsilon \ln \Trs{\exp(-\varepsilon\delta_A)(\op{\rho}_f) \op{\rho}_i}\big|_{\varepsilon=0}, \nonumber
\end{align}
which should be compared with \eqref{eq:realweakvalue}, \eqref{eq:imweakvalue}, and \eqref{eq:imweakvaluedef}.

We see that the imaginary part of the weak value can also be interpreted as half the logarithmic directional derivative of the pre-selection probability as the retrodictive state changes in the opposite direction along the flow generated by $\op{A}$.

\section{Examples}\label{sec:examples}
\subsection{Bohmian Mechanics}\label{sec:bohm}
To make the preceding abstract discussion of the weak value more concrete, let us consider a special case that has been recently discussed by \citet{Leavens2005}, \citet{Wiseman2007}, and \citet{Hiley2011}, where the operator $\op{A}=\op{p}$ being measured is the momentum operator of the system particle.  Since the wave-number operator $\op{k} = -\op{p}/\hbar$ generates a flow that is parametrized by the position $x$, then we expect from the discussion surrounding \eqref{eq:imweakvaluedef} that the imaginary part of a momentum weak value will give information about how the post-selection probability will change along changes in position.  

If we restrict our initial system state to be a pure state $\op{\rho}_i = \pr{\phi}$, and post-select the measurement of the momentum on a particular position $\op{P}_f = \pr{x}$, then the detector will have the linear response relations \eqref{eq:linear} with the complex weak value given by,
\begin{align}
  p_w &= \frac{\bra{x}\op{p}\ket{\phi}}{\ipr{x}{\phi}} = \frac{-i\hbar\partial_x \phi(x)}{\phi(x)}.
\end{align} 
We can split this value naturally into its real and imaginary parts by considering the polar decomposition of the initial system state $\phi(x) = r(x)\exp(iS(x))$,
\begin{align}\label{eq:bohmwv}
  p_w &= \hbar \partial_x S(x) - i \hbar \partial_x \ln r(x).
\end{align}

The real part of the weak value $\text{Re}\,p_w = \hbar \partial_x S(x)$ is the phase gradient, or \emph{Bohmian momentum} for the initial state, which we can now interpret operationally as the \emph{average momentum conditioned on the subsequent measurement of a particular} $x$ {in the ideal limit of no measurement disturbance}.  This connection between the real part of a weak value and the Bohmian momentum that was pointed out in \cite{Leavens2005,Wiseman2007} has recently allowed \citet{Kocsis2011} to experimentally reconstruct the averaged Bohmian trajectories in an optical two-slit interference experiment using such a von Neumann measurement.

The imaginary part of the weak value, $\text{Im}\,p_w = -\hbar\partial_x\ln r(x)$, on the other hand, is the logarithmic gradient of the root of the probability density $\rho(x) = |\phi(x)|^2 = r^2(x)$ for the particle at the point $x$.  Written in the form,
\begin{align}\label{eq:bohmimwv}
  2\text{Im}\,p_w = - \hbar \partial_x \ln \rho(x),
\end{align}
it describes the instantanous exponential rate of positional change of the probability density with respect to the particular post-selection point $x$, as expected.  This quantity, scaled by an inverse mass $1/m$, was introduced under the name ``osmotic velocity'' in the context of a stochastic interpretation of quantum mechanics developed by \citet{Nelson1966}, where it produced a diffusion term in the stochastic equations of motion for a classical point particle with diffusion coefficient $\hbar/2m$.  Nelson's interpretation was carefully contrasted with a stochastic interpretation for the Bohmian pilot wave by \citet{Bohm1989}, and the connection of the osmotic velocity with a weak value was recently emphasized by \citet{Hiley2011}.

Hence, the imaginary part of the momentum weak value does not provide information about a measurement of the momentum in the initial state.  Instead, it indicates the logarithmic directional derivative of the probability density for measuring $x$ along the flow generated by $\op{p}$.  The scaled derivative $-\hbar \partial_x$ appears since $\op{p} = -\hbar\op{k}$ and $\op{k}$ generates flow along the position $x$.  

\subsection{Qubit Observable}\label{sec:qubit}
To make the full von Neumann measurement process more concrete, let us also consider a simple example where $\op{A}$ operates on the two-dimensional Hilbert space of a qubit.  (See also \cite{Jozsa2007,DiLorenzo2008,Geszti2010,Wu2011,Haapasalo2011,Parks2011,Zhu2011,Kofman2011,Nakamura2011,Koike2011,DiLorenzo2011,Pan2011}.)  We can in such a case simplify the perturbative expansions \eqref{eq:opexpand} using the following identities,
\begin{subequations}\label{eq:qubit}
\begin{align}
  \op{A} &= A \op{\sigma}_3, \\
  \label{eq:qubitrhos}
  \op{\rho}_i &= \frac{1}{2}\left(\op{1}_s + \sum_k r_k \op{\sigma}_k \right), \\
  [\op{\sigma}_j,\op{\sigma}_k] &= 2 i\epsilon_{jkl}\op{\sigma}_l, \\
  \{\op{\sigma}_j,\op{\sigma}_k\} &= 2\delta_{jk}\op{1}_s,
\end{align}
\end{subequations}
where $\{\op{\sigma}_k\}_{k=1}^3$ are the usual Pauli operators, the components of the initial system state $\{r_k\}_{k=1}^3$ are real and satisfy the inequality $0\le\sum_k r_k^2\le1$, $\epsilon_{jkl}$ is the completely antisymmetric Levi-Civita pseudotensor, and $\delta_{jk}$ is the Kronecker delta.  We have defined $\op{\sigma}_3$ to be diagonal in the eigenbasis of $\op{A}$ and have rescaled the spectrum of $\op{A}$ for simplicity to zero out its maximally mixed mean $\Trs{\op{A} \op{1}_s/2} = 0$.  As a result, $\mean{A}_0 = A r_3$.

It follows that for positive integer $n$ the repeated actions of $\op{A}$ on the various qubit operators have the forms,
\begin{subequations}\label{eq:qubitidentitiesa}
\begin{align}
  (\text{ad}\op{A})^n(\op{1}_s) &= 0, \\
  (\text{ad}\op{A})^{2n-1}(\op{\sigma}_1) &= i\op{\sigma}_2 (2A)^{2n-1}, \\
  (\text{ad}\op{A})^{2n}(\op{\sigma}_1) &= \op{\sigma}_1 (2A)^{2n}, \\
  (\text{ad}\op{A})^{2n-1}(\op{\sigma}_2) &= -i\op{\sigma}_1 (2A)^{2n-1}, \\
  (\text{ad}\op{A})^{2n}(\op{\sigma}_2) &= \op{\sigma}_2 (2A)^{2n}, \\
  (\text{ad}\op{A})^n(\op{\sigma}_3) &= 0,
\end{align}
\end{subequations}
which collectively imply that,
\begin{subequations}\label{eq:qubitidentitiesb}
\begin{align}
  (\text{ad}\op{A})^{2n-1}(\op{\rho}_i) &= \frac{i}{2}(2A)^{2n-1}\left(r_1\op{\sigma}_2 - r_2\op{\sigma}_1\right), \\
  (\text{ad}\op{A})^{2n}(\op{\rho}_i) &= \frac{1}{2}(2A)^{2n}\left(r_1\op{\sigma}_1 + r_2\op{\sigma}_2\right),
\end{align}
\end{subequations}
and hence that the nonselective measurement operation has the exact form,
\begin{subequations}
\begin{align}
  \mathcal{E}(\op{\rho}_i) &= \op{\rho}_i + \frac{c(g)r_1 - s(g)r_2}{2}\op{\sigma}_1 \\
  &\qquad + \frac{c(g)r_2 + s(g) r_1}{2}\op{\sigma}_2, \nonumber \\
  c(g) &= \sum_{n=1}^\infty \frac{(-1)^n}{(2n)!}\left(\frac{2Ag}{\hbar}\right)^{2n}\mean{p^{2n}}_0, \\
  s(g) &= \sum_{n=1}^\infty \frac{(-1)^{n+1}}{(2n-1)!}\left(\frac{2Ag}{\hbar}\right)^{2n-1}\mean{p^{2n-1}}_0.
\end{align}
\end{subequations}
The correction term can be interpreted as a Rabi oscillation of the qubit that has been perturbed by the coupling to the detector.  Indeed, if the detector operator $\op{p}$ were replaced with a constant $p$, then the interaction Hamiltonian \eqref{eq:hamiltonian} would constitute an evolution term for the qubit that would induce Rabi oscillations around the $\op{\sigma}_3$ axis of the Bloch sphere, which would be the natural flow in state space generated by the action of $\op{A}$.  With the substitution $\op{p}\to p$ then $\mean{p^n}_0\to p^n$, so $c(g)\to \cos(2gAp/\hbar) - 1$ and $s(g) \to \sin(2gAp/\hbar)$, which restores the unperturbed Rabi oscillations.

Similarly, we find that the averaging operations for the detector position and momentum \eqref{eq:operxp} have the exact forms,
\begin{subequations}
\begin{align}
  \mathcal{X}(\op{\rho}_i) &= \mean{x}_0 \op{\rho}_i + \frac{c_x(g)r_1 - s_x(g)r_2}{2}\op{\sigma}_1 \\
  &\qquad + \frac{c_x(g)r_2 + s_x(g) r_1}{2}\op{\sigma}_2, \nonumber \\
  c_x(g) &= \sum_{n=1}^\infty \frac{(-1)^n}{(2n)!}\left(\frac{2Ag}{\hbar}\right)^{2n}\frac{\mean{\{p^{2n},x\}}_0}{2}, \\
  s_x(g) &= \sum_{n=1}^\infty \frac{(-1)^{n+1}}{(2n-1)!}\left(\frac{2Ag}{\hbar}\right)^{2n-1}\frac{\mean{\{p^{2n-1},x\}}_0}{2},
\end{align}
\end{subequations}
and,
\begin{subequations}
\begin{align}
  \mathcal{P}(\op{\rho}_i) &= \mean{p}_0 \op{\rho}_i + \frac{c_p(g)r_1 - s_p(g)r_2}{2}\op{\sigma}_1 \\
  &\qquad + \frac{c_p(g)r_2 + s_p(g) r_1}{2}\op{\sigma}_2, \nonumber \\
  c_p(g) &= \sum_{n=1}^\infty \frac{(-1)^n}{(2n)!}\left(\frac{2Ag}{\hbar}\right)^{2n}\mean{p^{2n+1}}_0, \\
  s_p(g) &= \sum_{n=1}^\infty \frac{(-1)^{n+1}}{(2n-1)!}\left(\frac{2Ag}{\hbar}\right)^{2n-1}\mean{p^{2n}}_0.
\end{align}
\end{subequations}
These operations differ from $\mathcal{E}$ only in how the various moments of the initial detector distribution weight the series for the Rabi oscillation.  In particular, given the substitutions $\op{p}\to p$ and $\op{x}\to x$, then $\mean{\{p^n,x\}/2}_0 \to p^n x$ and $\mean{p^{n+1}}_0 \to p^{n+1}$, so $c_x(g)\to x\, (\cos(2gAp/\hbar)-1)$, $s_x(g)\to x\, \sin(2gAp/\hbar)$, $c_p(g)\to p\,(\cos(2gAp/\hbar)-1)$, and $s_p(g)\to p\,\sin(2gAp/\hbar)$.  Therefore, if the detector remained uncorrelated with the system the averaging operations would reduce to $\mathcal{X}(\op{\rho}_i)\to x\, \mathcal{E}(\op{\rho}_i)$ and $\mathcal{P}(\op{\rho}_i)\to p\, \mathcal{E}(\op{\rho}_i)$, which are the decoupled intial detector means scaling the Rabi-oscillating qubit state.

Since we have assumed that $\op{A}$ does not have a component proportional to the identity, the symmetric product $\{\op{A},\op{\rho}_i\}/2 = A r_3 \op{1}_s / 2 = \mean{A}_0 (\op{1}_s/2)$ for a qubit will act effectively as an inner product that extracts the part of the initial state proportional to $\op{A}$.  Therefore, the correction to $\mean{A}_0$ in $\cmean{f}{x}_T$ that appears in \eqref{eq:xcorr} has the simple form,
\begin{align}\label{eq:qubita}
  g\frac{\Trs{\op{P}_f \mathcal{E}(\{\op{A},\op{\rho}_i\}/2)}}{\Trs{\op{P}_f\mathcal{E}(\op{\rho}_i)}} &= \frac{g\mean{A}_0}{\tilde{p}(f)}. 
\end{align}
where the conditioning factor,
\begin{align}
  \tilde{p}(f) &= 2 \Trs{\op{\rho}_f \mathcal{E}(\op{\rho}_i)}, \\
  &= 2 \Trs{\op{\rho}_f \op{\rho}_i} \nonumber \\
  &\qquad + (c(g)r_1 - s(g)r_2)\Trs{\op{\rho}_f \sigma_1} \nonumber \\
  &\qquad + (c(g)r_2 + s(g) r_1)\Trs{\op{\rho}_f \sigma_2}, \nonumber
\end{align}
is $(2/\Trs{\op{P}_f})$ times the total probability of obtaining the post-selection.  We have expressed $\tilde{p}(f)$ more compactly in terms of the retrodictive state \eqref{eq:retrodictivef} to show how the deviations from the initial state that are induced by $\op{A}$ become effectively averaged by the post-selection state.  In the absence of post-selection, the retrodictive state will be maximally mixed $\op{\rho}_f = \op{1}_s / 2$ and $\tilde{p}(f) \to 1$, recovering the unconditioned average $\mean{A}_0$.

The correction to the detector mean position $\mean{x}_0$ in \eqref{eq:xcorr} can be expressed in a similar way,
\begin{align}\label{eq:qubitx}
  \frac{\Trs{\op{P}_f \mathcal{X}(\op{\rho}_i)}}{\Trs{\op{P}_f \mathcal{E}(\op{\rho}_i)}} &= 
  \frac{1}{\tilde{p}(f)}\Big(2 \mean{x}_0 \Trs{\op{\rho}_f \op{\rho}_i} \\
  &\qquad + (c_x(g)r_1 - s_x(g)r_2)\Trs{\op{\rho}_f \op{\sigma}_1} \nonumber \\
  &\qquad + (c_x(g)r_2 + s_x(g)r_1)\Trs{\op{\rho}_f \op{\sigma}_2}\Big), \nonumber
\end{align}
as can the correction to the detector mean momentum $\mean{p}_0$ in \eqref{eq:pcorr},
\begin{align}\label{eq:qubitp}
  \frac{\Trs{\op{P}_f \mathcal{P}(\op{\rho}_i)}}{\Trs{\op{P}_f \mathcal{E}(\op{\rho}_i)}} &= 
  \frac{1}{\tilde{p}(f)}\Big(2 \mean{p}_0 \Trs{\op{\rho}_f \op{\rho}_i} \\
  &\qquad + (c_p(g)r_1 - s_p(g)r_2)\Trs{\op{\rho}_f \op{\sigma}_1} \nonumber \\
  &\qquad + (c_p(g)r_2 + s_p(g)r_1)\Trs{\op{\rho}_f \op{\sigma}_2}\Big). \nonumber
\end{align}

Expanding \eqref{eq:corrections} using \eqref{eq:qubita}, \eqref{eq:qubitx}, and \eqref{eq:qubitp} to linear order in $g$, we find the linear response \eqref{eq:linear} in terms of the real and imaginary parts of the qubit weak value,
\begin{subequations}\label{eq:qubitweakvalue}
\begin{align}
  \text{Re}A_w &= \frac{\mean{A}_0}{2\Trs{\op{\rho}_f \op{\rho}_i}}, \\
  2\text{Im}A_w &= \frac{\Trs{\op{\rho}_f \delta_A(\op{\rho}_i)}}{\Trs{\op{\rho}_f \op{\rho}_i}}, \\
  \delta_A(\op{\rho}_i) &= A(-r_2 \op{\sigma}_1 + r_1 \op{\sigma}_2).
\end{align}
\end{subequations}

\begin{figure}[t]
  \begin{center}
    \includegraphics[width=2.5in]{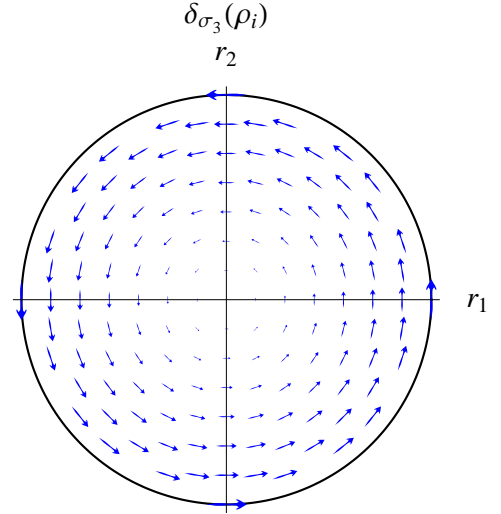}
  \end{center}
  \caption{(color online) The projection onto the plane $r_3 = 0$ of the qubit Bloch sphere, showing the vector field $\delta_{\sigma_3}(\op{\rho}_i) = -r_2 \op{\sigma}_1 + r_1 \op{\sigma}_2$ for arbitrary initial states $\op{\rho}_i = (\op{1} + r_1 \op{\sigma}_1 + r_2 \op{\sigma}_2 + r_3 \op{\sigma}_3)/2$.  The curves of the flow through this vector field are the Rabi oscillations around the $r_3$ axis that are generated by the unitary action of $\op{\sigma}_3$.  The quantity $2\text{Im}A_w$ \eqref{eq:qubitweakvalue} is the logarithmic rate of change of the post-selection probability \eqref{eq:imweakvaluedef} along this vector field.}
  \label{fig:blochfield}
\end{figure}

As expected, the real part contains information regarding the measurement of $\op{A}$ as an observable in the initial state, conditioned by the disturbance-free overlap between the predictive and retrodictive states.  The imaginary part, on the other hand, contains $\delta_A(\op{\rho}_i)$, which is a tangent vector field on the Bloch sphere---illustrated in Figure~\ref{fig:blochfield}---that corresponds to an infinitesimal portion of the Rabi oscillation being generated by $\op{A}$.  This tangent vector field contains only the components $r_1$ and $r_2$ from bases \emph{orthogonal} to $\op{A}$ in the initial state $\op{\rho}_i$, so $2\text{Im}A_w$ contains only the retrodictive averages of corrections to bases \emph{orthogonal} to $\op{A}$, and thus contains no information about the measurement of $\op{A}$ as an observable.  As discussed in \eqref{eq:imweakvaluedef}, $2\text{Im}A_w$ is the logarithmic rate of change of the post-selection probability along the vector field $\delta_A(\op{\rho}_i)$.  Scaling it by a small factor with units inverse to $A$ will produce a probability correction to linear order.  In the absence of post-selection, then $\op{\rho}_f \to \op{1}_s/2$, $\text{Re}A_w \to \mean{A}_0$, and $\text{Im}A_w \to 0$.

\subsection{Gaussian Detector}\label{sec:gaussian}
We can also apply our general results to the traditional case when the initial detector state in \eqref{eq:initcond} is a zero-mean Gaussian in position,
\begin{align}
  \ipr{x}{\psi} &= (2\pi \sigma^2)^{-1/4} \exp(-x^2 / 4 \sigma^2),
\end{align}
Then the measurement operators for position detection \eqref{eq:measopx} have the initial Gaussian form shifted by $g\op{A}$,
\begin{align}
  \op{M}_x &= \frac{1}{(2\pi \sigma^2)^{1/4}} e^{-(x - g \op{A})^2 / 4 \sigma^2}, 
\end{align}
while the conjugate measurement operators for momentum detection \eqref{eq:measopp} have the initial Gaussian modified by a unitary factor containing $\op{A}$,
\begin{align}
  \op{N}_p &= \left(\frac{2 \sigma^2}{\pi \hbar^2}\right)^{1/4}\, e^{-p^2 \sigma^2 / \hbar^2}\, e^{g p \op{A} / i \hbar}. 
\end{align}

The Wigner quasi-probability operator \eqref{eq:wigner} correspondingly decouples into a product of Gaussian distributions, with only the position shifted by the system operator,
\begin{align}
  \op{W}_{x,p} &= \frac{1}{\pi \hbar} e^{-(x-g\op{A})^2/2\sigma^2} e^{-2 p^2 \sigma^2 / \hbar^2}.
\end{align}

Marginalizing the Wigner operator over momentum and position separately produces the probability operators \eqref{eq:povmx} and \eqref{eq:povmp},
\begin{subequations}
\begin{align}
  \op{E}_x &= \frac{1}{\sqrt{2\pi \sigma^2}}\, e^{-(x-g\op{A})^2/2\sigma^2}, \\
  \op{F}_p &= \frac{\sigma}{\hbar}\sqrt{\frac{2}{\pi}}\, e^{-2p^2\sigma^2/\hbar^2}\, \op{1}_s.
\end{align}
\end{subequations}
As anticipated, the probability operator for momentum no longer contains any information about the system operator $\op{A}$ and is proportional to the identity, so measuring the momentum provides zero information about any system operator not proportional to the identity.

In the presence of post-selection we can also exactly compute the disturbance operations \eqref{eq:oper}, \eqref{eq:operx}, and \eqref{eq:operp} using the following identities for the Gaussian detector moments,
\begin{subequations}\label{eq:gaussianmoments}
\begin{align}
  \mean{p^{2n}}_0 &= \left(\frac{\hbar}{2\sigma}\right)^{2n}(2n-1)!!, \\
  \mean{p^{2n-1}}_0 &= 0, \\
  \mean{\{p^n,x\}/2}_0 &= 0, \\
  \frac{(2n-1)!!}{(2n)!} &= \frac{1}{2^n\, n!},
\end{align}
\end{subequations}
which hold for positive integer $n$.  We find the simple results,
\begin{subequations}\label{eq:gaussianops}
\begin{align}
  \mathcal{E}(\op{\rho}_i) &= \exp\left(-\frac{1}{2}\left(\frac{g}{2\sigma}\right)^2 (\text{ad}\op{A})^2\right)(\op{\rho}_i), \\
  \mathcal{X}(\op{\rho}_i) &= 0, \\
  \mathcal{P}(\op{\rho}_i) &= \frac{g}{i\hbar}\frac{\hbar^2}{4\sigma^2}(\text{ad}\op{A})(\mathcal{E}(\op{\rho}_i)).
\end{align}
\end{subequations}

The quantity $\epsilon = (g/2\sigma)^2$ with units inverse to $\op{A}^2$ emerges as the natural decoherence parameter, which we can see more clearly by rewriting the non-selective measurement operation in \eqref{eq:gaussianops} as,
\begin{subequations}\label{eq:lindblad}
\begin{align}
  \op{\rho}_i(\epsilon) &= \mathcal{E}(\op{\rho}_i) = \exp\left(\epsilon \mathcal{L}[\op{A}]\right)(\op{\rho}_i), \\
  \mathcal{L}[\op{A}](\op{\rho}_i) &= \op{A}\op{\rho}_i\op{A}^\dagger - \frac{1}{2}\{\op{\rho}_i,\op{A}^\dagger\op{A}\}.
\end{align}
\end{subequations}
The operation $\mathcal{L}[\op{A}](\op{\rho}_i)$ is the \emph{Lindblad operation} \cite{Lindblad1976,Breuer2007,Wiseman2009} that produces \emph{decoherence} in continuous dynamical systems, with $\op{A}$ playing the role of the Lindblad operator that decoheres the system.  Since $\partial_\epsilon\op{\rho}_i(\epsilon) = \mathcal{L}[\op{A}](\op{\rho}_i(\epsilon))$, the Gaussian measurement acts as an effective Lindblad evolution that \emph{decoheres} the system state with increasing $\epsilon$ via the action of $\op{A}$, but does not cause unitary disturbance along the natural flow of $\op{A}$ \footnote{That von Neumann coupling can lead to such Lindblad evolution in the case of \emph{continuous} indirect non-selective measurements with zero-mean detectors was also noted by \citet[\S 3.5.2, p.162]{Breuer2007}; however, here we make a single Gaussian detection after a duration of time, producing an effective flow parameter $(g/2\sigma)^2$.}.

The exact expressions for the conditioned Gaussian detector means follow from \eqref{eq:corrections} and \eqref{eq:gaussianops},
\begin{subequations}\label{eq:gaussianmeans}
\begin{align}
  \cmean{f}{x}_T &= g \frac{\Trs{\op{P}_f \mathcal{E}(\{\op{A},\op{\rho}_i\})}}{2\Trs{\op{P}_f \mathcal{E}(\op{\rho}_i)}}, \\
  \cmean{f}{p}_T &= \frac{g}{i\hbar}\frac{\hbar^2}{4\sigma^2} \frac{\Trs{\op{P}_f(\text{ad}\op{A})(\mathcal{E}(\op{\rho}_i))}}{\Trs{\op{P}_f \mathcal{E}(\op{\rho}_i)}}.
\end{align}
\end{subequations}
Surprisingly, the special properties of the Gaussian moments \eqref{eq:gaussianmoments} allow \eqref{eq:gaussianmeans} to be written in a form proportional to the real and imaginary part of a complex weak-value involving the \emph{decohered} system state \eqref{eq:lindblad} to all orders in the coupling strength $g$,
\begin{subequations}\label{eq:gaussianwv}
\begin{align}
  A_w(\epsilon) &= \frac{\Trs{\op{P}_f\op{A} \op{\rho}_i(\epsilon)}}{\Trs{\op{P}_f \op{\rho}_i(\epsilon)}}, \\
  \cmean{f}{x}_T &= g \text{Re}A_w(\epsilon), \\
  \cmean{f}{p}_T &= \frac{g}{\hbar} \frac{\hbar^2}{4\sigma^2} (2\text{Im}A_w(\epsilon)).
\end{align}
\end{subequations}
Following the interpretations outlined in this paper we can therefore understand the position shift $\text{Re}A_w(\epsilon)$ to all orders in $g$ as the \emph{average of the observable} $\op{A}$ \emph{in the decohered initial system state} $\op{\rho}_i(\epsilon)$ \emph{conditioned on the post-selection} $\op{P}_f$.  Similarly, we can understand the factor $2\text{Im}A_w(\epsilon)$ in the momentum shift to all orders in $g$ as the \emph{logarithmic directional derivative of the probability of post-selecting} $\op{P}_f$ \emph{given the decohered initial system state} $\op{\rho}_i(\epsilon)$ \emph{along the unitary flow generated by} $\op{A}$.

\begin{figure}[t]
  \begin{center}
    \includegraphics[width=3.5in]{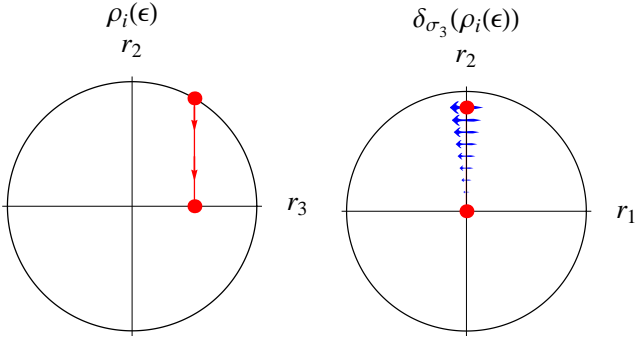}
  \end{center}
  \caption{(color online) Two projections of the Bloch sphere showing the pure decoherence of the specific state $\op{\rho}_i(\epsilon) = \exp(\epsilon \mathcal{L}[\op{\sigma}_3])(\op{\rho}_i) = (\op{1} + \exp(-2\epsilon) \sqrt{3}\op{\sigma}_2/2 + \op{\sigma}_3/2)/2$ due to the Gaussian detector \eqref{eq:qubitdecoheredstate}.  (left) The projection onto the plane $r_1 = 0$ showing the progressive collapse of $\op{\rho}_i(\epsilon)$ onto the $r_3$ axis with increasing $\epsilon$.  (right) The projection onto the plane $r_3 = 0$ showing the vector field $\delta_{\sigma_3}(\op{\rho}_i(\epsilon))$ during the progressive collapse.  Notably the quantity $2\text{Im}A_w(\epsilon)$ \eqref{eq:gaussianwv} is the rate of change of the post-selection probability \eqref{eq:imweakvaluedef} along this vector field for all $\epsilon$, but not along the purely decohering trajectory that $\op{\rho}_i(\epsilon)$ actually follows.}
  \label{fig:blochdissipate}
\end{figure}

If the measured operator is the qubit operator $\op{A} = A\op{\sigma}_3$ as in \eqref{eq:qubit}, then we can further simplify the expression \eqref{eq:gaussianops} using the identities \eqref{eq:qubitidentitiesb} to find,
\begin{align}\label{eq:qubitdecoheredstate}
  \mathcal{E}(\op{\rho}_i) &= \op{\rho}_i + \frac{1}{2}(e^{-(Ag/\sigma)^2/2}-1) \left(r_1\op{\sigma}_1 + r_2\op{\sigma}_2\right), \\
  &= \frac{1}{2}\left(\op{1} + r_3 \op{\sigma}_3 + e^{-(Ag/\sigma)^2/2}\left(r_1\op{\sigma}_1 + r_2\op{\sigma}_2\right)\right), \nonumber
\end{align}
which shows how the measurement decoheres the bases orthogonal to $\op{A}$ in the initial state with an increase in the dimensionless flow parameter $(Ag/\sigma)^2$ \footnote{The continuous measurement of a double quantum dot that is discussed in \cite{Williams2008,Dressel2010} can be mapped onto this problem so that $(Ag/\sigma)^2 \to \gamma t$ where $\gamma = 1/T_m$ is an inverse characteristic measurement time that acts as a dephasing rate due to the continuous non-selective measurement.  Hence, the results obtained therein are special cases of the exact solution \eqref{eq:gaussianqubitmeans}.}.  This decoherence is illustrated in Figure~\ref{fig:blochdissipate}.  The conditioned means \eqref{eq:gaussianmeans} of a Gaussian qubit detector consequently have the exact form,
\begin{subequations}\label{eq:gaussianqubitmeans}
\begin{align}
  \cmean{f}{x}_T &= g \frac{\mean{A}_0}{\tilde{p}(f)}, \\
  \cmean{f}{p}_T &= \frac{g}{\hbar}\frac{\hbar^2}{4\sigma^2}\frac{2A}{\tilde{p}(f)} e^{-(Ag/\sigma)^2/2}\, \times \\
  &\qquad\qquad (r_1 \Trs{\op{\rho}_f \op{\sigma}_2} - r_2 \Trs{\op{\rho}_f \op{\sigma}_1}), \nonumber \\
  \tilde{p}(f) &= 1 + r_3 \Trs{\op{\rho}_f \op{\sigma}_3} \\
  &\quad + e^{-(Ag/\sigma)^2/2}\left(r_1 \Trs{\op{\rho}_f\op{\sigma}_1} + r_2 \Trs{\op{\rho}_f\op{\sigma}_2}\right), \nonumber
\end{align}
\end{subequations}
to all orders in the coupling strength $g$.  When expanded to linear order in $g$, \eqref{eq:gaussianqubitmeans} reduces to \eqref{eq:linear} with the real and imaginary parts of the qubit weak value \eqref{eq:qubitweakvalue}, as expected.  

For contrast, as $g$ becomes large the unconditioned measurement of $\op{A}$ becomes essentially projective and the operation $\mathcal{E}$ almost completely decoheres the initial state \eqref{eq:qubitdecoheredstate} into the basis of $\op{A}$ as the pointer basis,
\begin{align}
  \mathcal{E}(\op{\rho}_i) &\approx \frac{1}{2}\left(\op{1}_s + r_3 \op{\sigma}_3\right). 
\end{align}
Hence, in this \emph{strong measurement} regime, the conditioned means \eqref{eq:gaussianqubitmeans} approximate,
\begin{subequations}\label{eq:gaussianqubitstrong}
\begin{align}
  \cmean{f}{x}_T &\approx g \frac{\mean{A}_0}{1 + r_3 \Trs{\op{\rho}_f \op{\sigma}_3}}, \\
  \cmean{f}{p}_T &\approx 0.
\end{align}
\end{subequations}
The position shift contains the average of $\op{A}$ in the \emph{decohered} initial system state $\mathcal{E}(\op{\rho}_i)$, conditioned by the post-selection.  Moreover, since the decohered initial system state $\mathcal{E}(\op{\rho}_i)$ is essentially diagonal in the basis of $\op{A}$, it will no longer Rabi oscillate, so the directional derivative along the flow generated by $\op{A}$ will be essentially zero.  Hence, the probability correction factor represented by $2\text{Im}A_w(\epsilon)$ vanishes.

\section{Conclusion}\label{sec:conclusion}

We have given an exact treatment of a conditioned von Neumann measurement for an arbitrary initial state and an arbitrary post-selection using the language of quantum operations.  The full form of the conditioned detector response \eqref{eq:corrections} naturally indicates how the measurement disturbance and conditioning from the post-selection modify the unconditioned detector response.  The corresponding linear response of the detector \eqref{eq:linear} can be parametrized by the \emph{generalized complex weak value} \eqref{eq:weakvalue}, but the origins of the real and imaginary parts differ.  

The real part of the weak value \eqref{eq:realweakvalue} stems directly from the measurement of $\op{A}$ made on the initial system state, and can be interpreted as the idealized zero-disturbance conditioned average of the operator $\op{A}$ acting in its role as an observable.  The imaginary part of the weak value \eqref{eq:imweakvalue}, on the other hand, contains no information about the measurement of $\op{A}$ as an observable, but instead arises from the \emph{disturbance} due to the von Neumann coupling.  We interpret it as the logarithmic directional derivative \eqref{eq:imweakvaluedef} of the post-selection probability along the unitary flow in state space generated by the operator $\op{A}$ in its role as the element of a Lie algebra.  The complex weak value therefore captures both halves of the dual role of the operator $\op{A}$ in the quantum formalism.

To illustrate this interpretation, we considered the weak value of momentum \eqref{eq:bohmwv} post-selected on a particular position.  Its real part is the Bohmian momentum representing the average momentum conditioned on a position detection, while its imaginary part \eqref{eq:bohmimwv} is proportional to the ``osmotic velocity'' that describes the logarithmic derivative of the probability density for measuring the particular position directed along the flow generated by the momentum.

Finally, we applied our exact solution to the useful special cases of a qubit operator and a Gaussian detector.  We showed how the natural qubit Rabi oscillations that would be generated by the von Neumann interaction Hamiltonian \eqref{eq:hamiltonian} become disrupted by the measurement.  We also showed that the Gaussian detector purely decoheres the initial system state into the basis of $\op{A}$ in the Lindblad sense, which allows the exact interpretation of the position and momentum shifts of the detector in terms of a complex weak value \eqref{eq:gaussianwv} that involves the \emph{decohered} system state.

\begin{acknowledgments}
This work was supported by the NSF Grant No. DMR-0844899, and ARO Grant No. W911NF-09-0-01417.
\end{acknowledgments}


%

\end{document}